\documentclass[fleqn,twoside]{article}
\usepackage{amsmath}
\usepackage{espcrc2}
\usepackage{graphicx}
\usepackage{slashed}

\newcommand{\Lag}{\mathcal{L}}   

\title{
{
\vspace{-3.0cm} \normalsize \hfill
\parbox{30mm}{HU-EP-04/61\\ SFB/CPP-04-55}
}\\[15mm]
Universality in the Gross-Neveu model}
\author{F. Knechtli\address[HUB]{Institut f\"ur Physik, 
        Humboldt-Universit\"at zu Berlin, Newtonstr. 15, 12489 Berlin, Germany},         
        T. Korzec\addressmark\thanks{presented by T. Korzec.},
        B. Leder\address[DESY]{DESY Zeuthen, Platanenallee 6, 15738 Zeuthen, Germany}, 
        U. Wolff\addressmark[HUB]}
       
\begin{document}

\begin{abstract}
We consider universal finite size effects in the large-$N$ limit of
the continuum Gross-Neveu model as well as in its discretized
versions with Wilson and with staggered fermions. After 
extrapolation to zero lattice spacing the lattice results are
compared to the continuum values. 
\vspace{1pc}
\end{abstract}

\maketitle

\section{Introduction}
The treatment of fermions in lattice simulations requires special care, and 
several different formulations have been found over the years. Since Aoki's
lattice-2000 plenary talk \cite{Aoki00} one may wonder, whether 
the models using Wilson-fermions and the staggered formulation 
really lie in the same universality class. It is difficult to clarify the 
situation in a four dimensional theory, where due to numerical costs the achievable
statistical errors are too high for precision checks. Hence we decided
to investigate a two dimensional toy model, which is also accessible to 
analytical calculations. 

For our study we choose the Gross-Neveu model, defined by the Euclidean action
density
\begin{equation}
   \Lag = \bar \psi_i \slashed \partial \psi_i - \frac{\lambda}{2N}(\bar\psi_i \psi_i)^2 \, .
\end{equation}
The fields $\psi_i$ are two component spinors and $i=1\ldots N$ is a flavor index.
The action possesses an $O(2N)$ flavor symmetry \cite{Dashen75}  and is invariant under
discrete chiral transformations
\begin{eqnarray}
   \psi_i        &\to& \gamma_5 \psi_i \nonumber \\
   \bar \psi_i   &\to& -\bar\psi_i\gamma_5 \, .
\end{eqnarray}
Spontaneous breakdown of this $\gamma_5$-symmetry leads to a dynamical mass generation \cite{Gross74}.
Moreover the model is renormalizable, asymptotically free and large-$N$ expandable. The large-$N$ limit
is taken keeping the dimensionless coupling constant $\lambda$ fixed.

An equivalent action, which is bilinear in the fermion fields results from the introduction
of an auxiliary scalar field $\sigma$
\begin{equation}
   \Lag^{\sigma} = \bar \psi_i( \slashed \partial+ \sigma  )\psi_i + \frac{N}{2\lambda}\sigma^2 \, .
\end{equation}

\section{Large-$N$ calculation}
In leading order of the large-$N$ expansion the fermion mass $m$ is given by the value of
the constant 
field $\sigma$ that satisfies the gap-equation
\begin{equation}
 \frac{\sigma}{\lambda} = \frac{1}{TL}{\rm tr}(\slashed \partial + \sigma)^{-1} \, ,
\end{equation}
where $L$ is the spatial and $T$ the temporal extent. In our notation $m(L)$ denotes
the finite volume fermion-mass\footnote{Note that in the $N\to\infty$ limit spontaneous symmetry
breaking can occur even in finite volume.}, and $m(\infty) \equiv m$.
We are interested in the universal curve $m(L)L$ versus $mL$, similar to the 
L\"uscher-Weisz-Wolff-coupling \cite{Luscher91}.
\subsection{Continuum}
\begin{figure} [htb]      
      \vspace{-0.5 cm}

      \includegraphics [width = \linewidth] {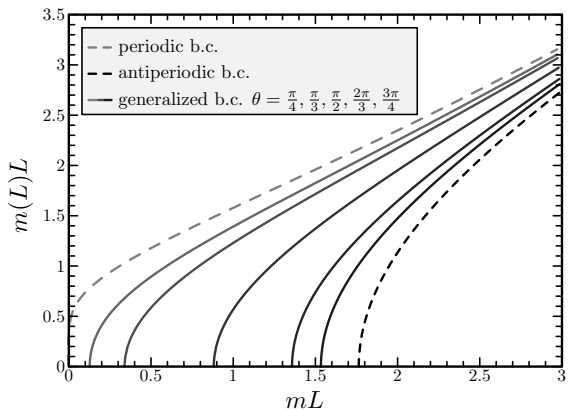} 
      
      \vspace{-0.5cm}
      
      \caption{Universal curves $m(L)L$ versus $mL$ obtained from solutions of the gap-equation of
               the continuum Gross-Neveu model.}
      \label{continuum}
\end{figure}

In the continuum theory the calculation can be performed in analogy to the
finite temperature calculation \cite{Wolff85}. In infinite volume the gap-equation can
be solved in closed form. It has solutions at $\sigma = 0$ (maximum of the effective potential)
as well as at $\sigma = \pm m$ (minima) with
\begin{equation}\label{infvolsol}
   m = \frac{\Lambda}{\sinh\frac{\pi}{\lambda}} \, ,
\end{equation}
where $\Lambda$ is a cutoff on the spatial momentum. 
In finite volume ($L$ finite, $T\to \infty$) we have to specify the spatial boundary conditions. 
Our choice are b.c.
of the form $\psi(x_0,x_1+L) = e^{i\theta} \psi(x_0,x_1)$ which include periodic and antiperiodic b.c. as
special cases.
Eq.~(\ref{infvolsol}) can be used
to eliminate the bare parameter $\lambda$ in the finite volume gap-equation. The
physical scale in the renormalized theory is given by $m$, which is kept constant while the cutoff is removed,
allowing us to find solutions
$m(L)L$ versus $mL$. Fig. \ref{continuum} shows the results which strongly 
depend on the chosen angle $\theta$. For all $-\pi<\theta<\pi$ different from zero there is a ``critical volume'' below 
which no spontaneous symmetry breaking takes place
\begin{equation*}
    L_c=\frac{4\pi}{m}\exp\left[-\frac{\pi}{|\theta|}-\gamma - \sum_{n=1}^\infty 
         \left(\frac{\theta}{2\pi}\right)^{2n}\zeta(2n+1)  \right] \, , 
\end{equation*}
where $\gamma$ is the Euler-Mascheroni-constant and $\zeta$ the Riemann-zeta-function.
\subsection{Wilson fermions}
\begin{figure} [htb]      
      \vspace{-0.5 cm}

      \includegraphics [width = \linewidth] {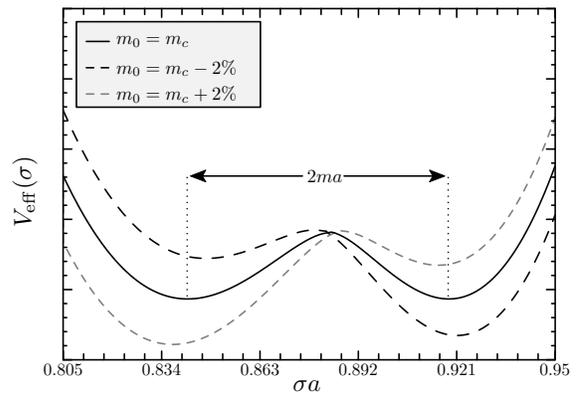} 
      
      \vspace{-0.5cm}
      
      \caption{The effective potential with Wilson fermions for some values of $m_0$ around 
      the critical one. $L/a = T/a = \infty$.}
      \label{veffwilson}
\end{figure}

\begin{figure} [htb]      
\vspace{-0.5 cm}

      \includegraphics [width = \linewidth] {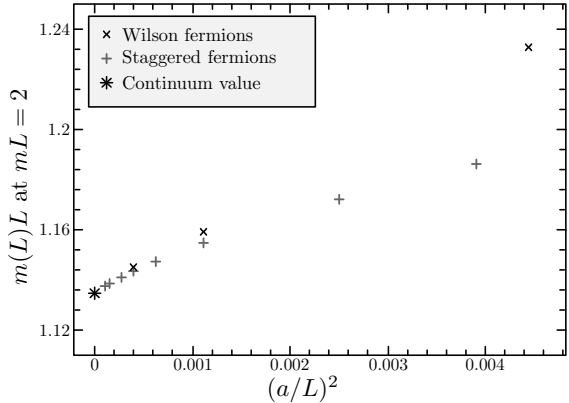} 
      
      \vspace{-0.5cm}
      
      \caption{Approach of $m(L)L$ to the continuum value at $mL = 2$ for staggered- and Wilson-fermions.
               Antiperiodic boundary conditions.}
      \label{aproach}
\end{figure}

The Wilson term breaks chiral symmetry explicitly which can lead to difficulties.
The Gross-Neveu model with Wilson fermions was pioneered by Aoki and Higashijima \cite{Aoki85}.
The authors found that in order to recover the $\gamma_5$-symmetry in the continuum limit, 
it is necessary to introduce a bare mass parameter $m_0$ and tune it to a critical value.
The fermion mass $m$ can be
defined as half of the distance between the two degenerate (at critical $m_0$) minima of 
the effective potential
\begin{equation}
   V_{\rm eff} = \frac{\sigma^2}{2\lambda} - \frac{1}{TL} {\rm tr} \ln(D_W+\sigma+m_0)\, ,
\end{equation}
which is not symmetric at finite lattice spacing~$a$ (Fig. \ref{veffwilson}). $D_W$ is the free 
Wilson-Dirac operator. This definition of $m$
coincides with that in the continuum theory for $a \to 0$. 
From fermion masses in finite and infinite volume we can construct curves corresponding to 
those in Fig.~\ref{continuum} for a series of lattice spacings.
The approach
to the continuum for a fixed value of $mL$ is shown in Fig.~\ref{aproach}.

\subsection{Staggered fermions}
With  Kogut-Susskind fermions it first has to be
clarified how to treat the four fermion interaction.
One method is to perform the spin diagonalization within
the free theory, keep only one component of the fields ($\bar\chi,\chi$) and construct taste fields
($\bar q, q$)
from one-component fields within a hypercube (a square in 2d). Once
the taste fields are constructed, 
the four-fermion interaction $(\bar q q)^2$ can be written down and
expressed in terms of the one-component fields,
which leads to the action
\begin{eqnarray}
   S   &=&\sum_x \bar \chi_i(x) \eta_\mu \tilde\partial_\mu \chi_i(x)  \\
       &-& \frac{\lambda}{N} \sum_X \left(\sum_\rho \bar\chi_i(2X+\rho)\chi_i(2X+\rho)\right)^2 \, . \nonumber
\end{eqnarray}
$X$ denotes the hypercubes and $\rho$ the position within a hypercube, such that $x = 2X+\rho$.
In two dimensions the phases are given by $\eta_\mu = (-1)^{\mu x_0}$. 
The symbol $\tilde \partial_\mu$ stands for the symmetric lattice derivative.
Again the four-fermion interaction term can be traded for a scalar field.
In \cite{Jolicoeur86} it was pointed out, that the lack of translation symmetry
by one lattice spacing complicates a perturbative treatment of this model.
An alternative staggered action, which cures this problem has got the interaction term
\begin{equation}
   \frac{\lambda}{4N} \sum_x \left(\sum_\rho \bar\chi_i(x+\rho)\chi_i(x+\rho)\right)^2 
\end{equation}
Both formulations lead to the gap-equation
\begin{equation}
   \frac{2\sigma}{\lambda} = \frac{1}{TL} {\rm tr}(\eta_\mu \tilde\partial_\mu +\sigma)^{-1}\, .
\end{equation}
Numerical solutions at a fixed $mL$ for different lattice spacings are shown in 
Fig.~\ref{aproach}.

\section{Conclusions}
Our large-$N$ calculation in the Gross-Neveu model allows for the following conclusions:
\begin{itemize}
   \item Both the naive staggered and the Wilson formulation lead to the correct 
         continuum limit of the finite volume massgap.
   \item As in QCD, with Wilson fermions an additive mass renormalization is necessary,
         which is absent with staggered fermions.
   \item With staggered fermions the construction of physical fields is more
         involved and special care has to be taken when interactions are introduced.	 	 
\end{itemize}
It should be mentioned however, that the leading order of the large-$N$ expansion
is insensitive to problems that might occur at finite $N$. Therefore a 
calculation of the $1/N$-correction would be desirable.
In a Monte-Carlo study at finite $N$ some additional problems
will show up, e.g. the lack of true spontaneous symmetry breaking in finite volume at finite $N$ and the
problem of determining the critical bare mass with Wilson fermions.

Eventually we plan also to investigate the taste reduction by taking fractional powers of
the staggered determinant. There the question is open, whether a local fermion operator 
can be identified to give a solid basis to this approach \cite{Bunk04}.

\end{document}